\documentclass[12pt]{iopart}

\begin{document}

\title{Complete Solution of the Kinetics in a Far-from-equilibrium Ising Chain}
\author{M. Mobilia, R.K.P. Zia and B. Schmittmann}

\address{Department of Physics and\\
Center for Stochastic Processes in Science and Engineering,\\
Virginia Tech, Blacksburg, VA 24061-0435, USA}
\date{May 14, 2004}

\begin{abstract}
The one-dimensional Ising model is easily generalized to a \textit{genuinely
nonequilibrium} system by coupling alternating spins to two thermal baths at
different temperatures. Here, we investigate the full time dependence of
this system. In particular, we obtain the evolution of the magnetisation,
starting with arbitrary initial conditions. For slightly less general
initial conditions, we compute the time dependence of all correlation
functions, and so, the probability distribution. Novel properties, such as
oscillatory decays into the steady state, are presented. Finally, we comment
on the relationship to a reaction-diffusion model with pair annihilation and
creation.
\end{abstract}

\maketitle
\date{May, 14 2004}

\textit{Introduction:} With their connections to both fundamental issues of
statistical mechanics and applications to a range of disciplines,
nonequilibrium many-body systems have received much attention recently (see
e.g. \cite{Privman} and references therein). Despite these efforts a
comprehensive theoretical approach is still lacking: As yet, there is no
equivalent of the Gibbs ensemble theory for nonequilibrium systems.
Consequently, much of the observed macroscopic properties of such systems
are sensitive to the underlying microscopic dynamics, in contrast to systems
in thermal equilibrium. In particular, most progress in this field is made
by studying paradigmatic models \cite{Privman}, with a master equation
governing their evolution. In this context, exact solution of simple models
are very valuable (but also very rare), as they can be used as milestones to
develop approximate/numerical schemes and to shed light on some general
properties of related models. The Ising model is a good example with a
venerable history \cite{Privman,Glauber}. Recently, an interesting
generalization of it was studied \cite{RZ}: a kinetic Ising chain in which
spins at alternating sites are coupled to thermal baths of two\textit{\
different}\emph{\ }temperatures. As a result, at long times, this system
reaches a stationary state with inherently \textit{nonequilibrium}
properties, e.g., a non-zero heat flux through the system\cite{RZ}.
Subsequently, \textit{all }correlation functions were computed exactly, so
that the full stationary distribution is known \cite{Beate1}. Various other
versions of two- or multiple-temperature models have been studied \cite
{TT}.

In this letter we show that the \textit{dynamic }aspects of this model are
also accessible. As illustrations, we present the complete solution of the
time-dependent magnetization and two-spin correlation function. These
quantities can then be exploited to compute \textit{all} other correlation
functions, so that the time-dependent probability distribution is also (at
least formally) known. The key behind the solvability of this system lies in
the simple structure of its Glauber-like kinetics, so that the usual BBGKY
hierarchy \cite{BBGKY} decomposes into a closed set of \textit{linear}
equations for each $N$-spin correlation function.

\textit{Model specifications: }We consider an Ising spin chain defined on a
ring of $L$ sites. For simplicity, we choose $L$ to be even and denote a
spin at site $j$ by $\sigma _{j}$ (which assumes values $\pm 1$). The spins
interact ferromagnetically via the usual nearest-neighbor Hamiltonian: $%
\mathcal{H}=-J\sum_{j}\sigma _{j}\sigma _{j+1}$ ($J>0$; the
anti-ferromagnetic case can be accessed by a gauge transformation). Next, we endow the system with a
Glauber-like dynamics, but couple spins on the even and odd sites to
reservoirs at temperatures $T_{e}$ and $T_{o}$, respectively. For $T_{e}\neq
T_{o}$, this dynamics violates detailed balance \cite{RZ,Beate1,Beate2} and
leads to a nonequilibrium stationary (but probably non-Gibbsian) state.
Denoting a configuration ($\sigma _{1},\sigma _{2},\dots ,\sigma _{L}$) of
our system by $\{\sigma \}$, we implement this dynamics through a master
equation for time-dependent probability distribution $P(\{\sigma \},t)$: 
\begin{equation}
\partial _{t}P(\{\tilde{\sigma}\},t)=\sum_{\{\sigma \}}\left[ W\left( \{%
\tilde{\sigma}\};\{\sigma \}\right) P(\{\sigma \},t)-W\left( \{\sigma \};\{%
\tilde{\sigma}\}\right) P(\{\tilde{\sigma}\},t)\right]  \label{master}
\end{equation}
with transition rates 
\begin{equation}
\label{SF}
W\left( \{\tilde{\sigma}\};\{\sigma \}\right) =\sum_{j}\frac{1}{2}\left[
1-\gamma _{j}\tilde{\sigma}_{j}\left( \frac{\sigma _{j-1}+\sigma _{j+1}}{2}%
\right) \right] \prod_{k\neq j}\delta \left( \tilde{\sigma}_{k},\sigma
_{k}\right) \,\,. 
\end{equation}
Here, $\gamma _{j}$ is $\gamma _{e}\equiv \mathrm{tanh}\left(
2J/k_{b}T_{e}\right) $ for even $j$ and $\gamma _{o}\equiv \mathrm{tanh}%
\left( 2J/k_{b}T_{o}\right) $ for odd $j.$ For convenience, the overall
factor of $1/2$ is chosen so that all decays follow a simple $e^{-t}$-law in
the $J=0$ limit.

Our goal is to compute \textit{all} correlation functions $\langle \sigma
_{j_{1}}\dots \sigma _{j_{n}}\rangle _{t}\equiv \sum_{\{\sigma \}}\sigma
_{j_{1}}\dots \\ \dots \sigma _{j_{n}}P(\{\sigma \},t)$ and to represent the complete
solution for $P(\{\sigma \},t)$ by the relation \cite{Glauber}: 
\begin{equation}
\label{rel}
2^{L}P(\{\sigma \},t) = 1+\sum_{i}\sigma _{i}\langle \sigma _{i}\rangle
_{t}+\sum_{j<k}\sigma _{j}\sigma _{k}\langle \sigma _{j}\sigma _{k}\rangle
_{t}  \nonumber  \label{rel}+\sum_{j<k<l}\sigma _{j}\sigma _{k}\sigma _{l}\langle \sigma _{j}\sigma
_{k}\sigma _{l}\rangle _{t}+\dots 
\end{equation}
Recently, the stationary distribution, $P(\{\sigma \},t=\infty )$, was found
in this manner \cite{Beate1}. We will first present the solutions for $%
\langle \sigma _{i}\rangle _{t}$ and $\langle \sigma _{j}\sigma _{k}\rangle
_{t}$ and then, in terms of these, provide expressions for the other
correlations.

For Ising chains in contact with only one thermal bath, it is well known
that a gauge transformation (changing the sign of every other spin) relates
a system coupled to $T<0$ to one coupled to $T>0$. Here, it is clear that
such a transformation is applicable if the signs of \emph{both} $T_e$ and $%
T_o$ are changed. Thus, we will investigate explicitly two cases: one when
both $T$'s are positive and the other, when they are of opposite signs. As
noted in \cite{RZ}, quite unusual properties arise in the latter case. Here,
we will find similar oscillatory behaviour, but in the time domain.

\vspace{0.2cm}

\textit{The time-dependent magnetization:} The single-spin function $\langle
\sigma _{j}\rangle _{t}$ is, of course, just the $t$-dependent magnetisation
at site $j$, which we denote by $m_{j}(t)$. With the master equation (\ref
{master}), the equation of motion of the local magnetisation reads 
\begin{equation}
\frac{d}{dt}m_{j}(t)=\frac{\gamma _{j}}{2}\left[
m_{j-1}(t)+m_{j+1}(t)\right] -m_{j}(t).  \label{MSe}
\end{equation}
Given any initial $m_{j}(0)$, the full $t$-dependent magnetisation is just 
\[
m_{j}(t)=\sum_{k}M_{jk}\left( t\right) m_{k}(0)\,\,, 
\]
where $M_{jk}\left( t\right) $ is the ``propagator''.
In equilibrium ($\gamma _{j}=\gamma $), Glauber \cite{Glauber} obtained
 $M_{jk}\left( t\right)=e^{-t}I_{k-j}(\gamma t)$ , where $I_{n}(t)$ denotes usual
modified Bessel function \cite{Abramowitz}. 
Though our system is not in
equilibrium, we exploit this result by defining a modified 
magnetisation, $m_{j}\left( t\right) /\sqrt{\gamma
_{j}}$, which allows to reduce (\ref{MSe}) to the Glauber case with $\gamma $
replaced by 
$\alpha \equiv \sqrt{\gamma _{e}\gamma _{o}},\,$
the same parameter that enters into the steady-state correlations functions
in \cite{RZ,Beate1}. In other words, provided $T_{e},T_{o}>0$, we can
associate our system with an equilibrium one, coupled to a bath with
temperature $T_{{\rm eff}}$, given by 
$\tanh \left[ 2J/k_{b}T_{{\rm eff}}\right] =\sqrt{\tanh \left[ 2J/k_{b}T_{e}\right]
\tanh \left[ 2J/k_{b}T_{o}\right] }$.
To be precise, we have 
\begin{equation}
M_{jk}\left( t\right) =e^{-t}\sqrt{\frac{\gamma _{j}}{\gamma _{k}}}%
I_{k-j}(\alpha t)\;\; ; \;\; \alpha \equiv \sqrt{\gamma _{e}\gamma _{o}},\, \label{M}
\end{equation}
which indicates that $m_{j}(t)$ suffers (linear combinations of) exponential
decay similar to the equilibrium case. The more interesting case involves
baths of\textit{\ opposite} signs. Then, we can either rely on analytic
continuation of $I_{n}(\alpha t)$ to pure imaginary $\alpha $ or solve
equation (\ref{MSe}) explicitly. The result involves oscillations with a
simple exponential envelope: $e^{-t}$. As an illustration, if the initial
magnetisation is homogeneously $\bar{m}$, then

\[
m_{j}(t)=\bar{m}e^{-t}\left[ \cos (\left| \alpha \right| t)+\frac{\gamma _{j}%
}{|\alpha |}\sin (\left| \alpha \right| t)\right] 
\]
Interestingly, the frequency of the oscillations increases as the $T$'s are
lowered. Such remarkable properties can perhaps be traced to a mild form of
``frustration'', arising from the competition between the two baths. While
the effects of the positive $T$ reservoir is to align spins with its
neighbours, the other bath struggles to ``anti-align'' them. Other notable
behaviours occur at the limits. If, say, $T_{o}\rightarrow \infty $ ($\gamma
_{o}\rightarrow 0$), then the spins at the odd sites decouple and $
m_{2j+1}(t)$ decays purely by $e^{-t}$. This allows us to integrate (\ref
{MSe}) for the even sites: $m_{2j}(t)=m_{2j}(0)\,e^{-t}+\gamma
_{e}t\,e^{-t}\,\left\{ m_{2j-1}(0)+m_{2j+1}(0)\right\} /2$. At first sight,
it may seem surprising that the effects of the neighbours linger longer.
However, this aspect is due entirely to the details of the dynamics here
(random sequential update based on the average spins of the neighbours). At
the other extreme, there is qualitatively new behaviour only when \emph{both}
$T$'s vanish. As expected, the uniform component of the initial
magnetisation survives. (As a reminder, note that if the $T$'s $\rightarrow
0_{-}$, only the staggered component remains.)

\vspace{0.2cm}

\textit{Equal-time correlations: }Next we turn to the time-dependent
two-point correlation function: $\langle \sigma _{j}\sigma _{k}\rangle
_{t}-\langle \sigma _{j}\rangle _{t}\langle \sigma _{k}\rangle _{t}$. In
most studies of the Ising chain, the second term is typically neglected,
since there is generally no spontaneous magnetisation. Of course, we have
the result for $m_{j}\left( t\right) $ from above and will focus only on $%
\langle \sigma _{j}\sigma _{k}\rangle _{t}$. The transformation
above can still be exploited here, but it is not compatible with the 
``boundary condition'' $\langle \sigma _{k}\sigma
_{k}\rangle _{t}\equiv 1$. 
Nevertheless, we are able to use the method of
images \cite{Glauber} to find the general solution, which is rather
involved and will be presented elsewhere \cite{prep}. Here, let us
illustrate the results by restricting ourselves to a simpler case, namely,
one with (period-2) translationally invariance (as in \cite{RZ}). Then, we
need to consider only four functions (of one variable: $k-j$), namely, the
correlation between spins at two even sites, two odd sites, and one of each.
We denote these by $c_{2n}^{ee}(t)\equiv \langle \sigma _{2\ell }\sigma
_{2(\ell +n)}\rangle _{t}$, $c_{2n}^{oo}(t)\equiv \langle \sigma _{2\ell
-1}\sigma _{2\ell -1+2n}\rangle _{t}$, $c_{2n-1}^{eo}(t)\equiv \langle
\sigma _{2\ell }\sigma _{2\ell +2n-1}\rangle _{t}$ and $c_{2n-1}^{oe}(t)
\equiv \langle \sigma _{2\ell +1}\sigma _{2\ell +2n}\rangle _{t}$. Of
course, $\langle \sigma _{j}\sigma _{k}\rangle _{t}=\langle \sigma
_{k}\sigma _{j}\rangle _{t}$, so that the first pair are even in $n$, and
the last two are related by $c_{-2n+1}^{eo}(t)=c_{2n-1}^{oe}(t)$. Thus,
there is no need to study $n<0$ cases. Finally, we have the boundary
condition (BC): $c_{0}^{ee}=c_{0}^{oo}=1$, the main source of complication
here in comparison with the analysis for $m_{j}\left( t\right) $.

From the master equation (\ref{master}), we find that they satisfy (for $n>0$%
) 
\begin{eqnarray}
\frac d{dt}c_{2n}^{ee} &=&-2c_{2n}^{ee}+\frac{\gamma _e}%
2[c_{2n-1}^{oe}+c_{2n+1}^{oe}+c_{2n-1}^{eo}+c_{2n+1}^{eo}]\,\,,  \label{cor1}
\\
\frac d{dt}c_{2n}^{oo} &=&-2c_{2n}^{oo}+\frac{\gamma _o}%
2[c_{2n-1}^{eo}+c_{2n+1}^{eo}+c_{2n-1}^{oe}+c_{2n+1}^{oe}]\,\,,  \label{cor2}
\\
\frac d{dt}c_{2n-1}^{eo} &=&-2c_{2n-1}^{eo}+\frac{\gamma _e}%
2[c_{2n}^{oo}+c_{2n-2}^{oo}]+\frac{\gamma _o}2[c_{2n}^{ee}+c_{2n-2}^{ee}]\,%
\,,  \label{cor3} \\
\frac d{dt}c_{2n-1}^{oe} &=&-2c_{2n-1}^{oe}+\frac{\gamma _o}%
2[c_{2n}^{ee}+c_{2n-2}^{ee}]+\frac{\gamma _e}2[c_{2n}^{oo}+c_{2n-2}^{oo}]\,%
\,.  \label{cor4}
\end{eqnarray}
These simplify considerably, since the combinations $\gamma
_ec_{2n}^{oo}-\gamma _oc_{2n}^{ee}$ and $c_{2n-1}^{eo}-c_{2n-1}^{oe}$ decouple and
just decay with $e^{-2t}$ from their initial values. Meanwhile, the other
combinations, $\gamma _ec_{2n}^{oo}+\gamma _oc_{2n}^{ee}$ and $%
c_{2n-1}^{eo}+c_{2n-1}^{oe}$ are coupled, but the quantities ($n\geq 0$) 
\begin{eqnarray}
a_{2n}\left( t\right)  \equiv \frac 12\left[ \gamma _ec_{2n}^{oo}\left(
t\right) +\gamma _oc_{2n}^{ee}\left( t\right) \right] \,\,; \,\, 
a_{2n-1}\left( t\right)  \equiv \frac \alpha 2\left[ c_{2n-1}^{eo}\left(
t\right) +c_{2n-1}^{oe}\left( t\right) \right]   \label{c-def},
\end{eqnarray}
 allow to reduce equations (\ref{cor1}-\ref{cor4}) to a single equation: 
\begin{equation}
\frac d{dt}a_j=-2a_j+\alpha [a_{j-1}+a_{j+1}]\,\,,\quad j>0\,\,,  \label{da}
\end{equation}
with the BC 
\begin{equation}
a_0\left( t\right) =\bar{\gamma}\,;\quad \bar{\gamma}\equiv \left( \gamma
_e+\gamma _o\right) /2\,\,.  \label{a0}
\end{equation}
Now, equations (\ref{da},\ref{a0}) are precisely those encountered by
Glauber \cite{Glauber}, the only differences being $\alpha ,\bar{\gamma}$
instead of $\gamma ,1$. An immediate consequence is the steady-state result,
which takes the form 
$a_k\left( t\rightarrow \infty \right) =\bar{\gamma}\omega _0^k\,;\quad
\omega _0\equiv \tanh \left[ J/k_bT_{eff}\right] \,\,,$
in agreement with those in Refs. \cite{RZ,Beate1} ($\lambda $ in \cite{RZ} = 
$\omega _0^2$ here). As for the complete solution with arbitrary initial
correlations $\langle \sigma _j\sigma _k\rangle _0$, we could simply rewrite
Glauber's solution here. Instead, let us illustrate how to derive a more
compact form for the time dependence, in a simple example: $\langle \sigma
_j\sigma _{k\neq j}\rangle _0\equiv 0$ ({\it i.e.}, uncorrelated initial spins if $%
m_j\left( 0\right) =0$ also). Then, $\gamma _ec_{2n}^{oo}-\gamma _oc_{2n}^{ee}$
and $c_{2n-1}^{eo}-c_{2n-1}^{oe}$ ($n>0$) simply remain zero for all time,
so that $c_{2n}^{ee,oo}=a_{2n}/\gamma _{o,e}$ and $%
c_{2n-1}^{eo,oe}=a_{2n-1}/\alpha $. To see how $a_k$ builds up to the
steady-state value, we exploit the Laplace transforms: $\hat{a}_k(s)\equiv
\int e^{-st}\,a_k(t)$. Condition (\ref{a0}) leads to $\hat{a}_0(s)=
\bar{\gamma}/s\,\,, $
while equations (\ref{da}) can be solved by an Ansatz similar to the one in
Ref.\cite{RZ,Beate1}: 
$\hat{a}_k(s)=A(s)\omega (s)^k\,\,,\,\,\,k>0$. Inserting 
these into (\ref{da}), we find ($A\neq 0$)
$\omega ^{-1}+\omega =\left( 2+s\right) /\alpha $. 
Meanwhile, $\hat{a}_0(s)$ leads to $A(s)=\bar{\gamma}/s $.
As expected, we need only $\omega _0\equiv \omega \left( s=0\right) $ for
the steady state, since the singularities of $\omega $ lie at $s\leq
-2\left( 1-\alpha \right) $ and the pole in $A$ controls the $t\rightarrow
\infty $ limit. For finite $t$, using properties of Laplace transforms
\cite{Abramowitz} we arrive at a simple result:
\begin{equation}
\langle \sigma _j\sigma _{k\neq j}\rangle _t=\frac{\bar{\gamma}}{\alpha ^2}%
\sqrt{\gamma _j\gamma _k}\left| j-k\right| \int_0^{2t}\frac{d\tau }\tau
\,e^{-\tau }I_{\left| j-k\right| }(\alpha \tau )\,\,.  \label{ETcor}
\end{equation}

With this expression, we can study long-time behaviors of these
correlations. Like the case for $m_{j}\left( t\right) $, the leading decay
(towards their steady-state values) is monotonic: $t^{-3/2}e^{-2(1-\alpha
)t} $. As in the case for $m\left( t\right) $, if $T_{e}T_{o}<0$, $\alpha $
turns pure imaginary, and we find\textit{\ }oscillatory behavior, damped by $%
t^{-3/2}e^{-2t}$ \cite{prep}. Other unusual properties emerge at certain
limits. If, for example, $T_{o}\rightarrow \infty $ ($T_{e}$ finite), then $%
\alpha ,\gamma _{o}\rightarrow 0$. Starting with our initial condition, only
the nearest and next-nearest neighbour correlations will build up to
non-vanishing values \cite{Beate1}: $c_{1}(t)=\left( \gamma _{e}/4\right)
\left( 1-e^{-2t}\right) $ and $c_{2}^{ee}(t)=\left( \gamma _{e}^{2}/8\right)
\left[ 1-e^{-2t}\left( 1+2t\right) \right] $ ($c_{2}^{oo}(t)=0$, of course).
A curious limit is $T_{e}=-T_{o}$ in which $\bar{\gamma}=0$, so that $%
\langle \sigma _{j}\sigma _{k\neq j}\rangle _{\infty }\equiv 0$. Thus, an
initially uncorrelated state \emph{never} succeeds in building correlations.
We caution that this is a somewhat singular example, as initital
correlations are expected to survive with $e^{-2t}$ tails. Finally, we may
consider the most extreme case: $T_{e,o}\rightarrow 0_{+,-}$. Assuming their
magnitudes are unequal in the limiting process, then all correlations are
suppressed by $\bar{\gamma}=O\left( \exp \left\{ -2J/k_{b}T_{>}\right\}
\right) $, where $T_{>}$ is the bigger of $T_{e}$, $\left| T_{o}\right| $.

\vspace{0.2cm}

\textit{General two-point correlation functions:} Next, let us ask how a
spin on site $k$ at time $t$ is correlated with the spin on site $j$ at a
later time $t+t^{\prime }$. Following Glauber \cite{Glauber} again, we
define 
\begin{equation}
c_{j,k}(t^{\prime };t)\equiv \sum_{\{\sigma ^{\prime }\},\{\sigma \}}\sigma
_{j}^{\prime }\sigma _{k}\mathcal{P}(\{\sigma ^{\prime }\},t+t^{\prime
}|\{\sigma \},t)P(\{\sigma \},t)
\end{equation}
where $\mathcal{P}(\{\sigma ^{\prime }\},t+t^{\prime }|\{\sigma \},t)$, is
the probability to find the system with configuration $\{\sigma ^{\prime }\}$
at time $t+t^{\prime }$ \emph{conditioned} on the configuration being $%
\{\sigma \}$ at $t$. Being the propagator for the entire system, $\mathcal{P}
$ can be represented as a sum of terms, each of which involves the evolution
of $N$-spin functions: $\langle \sigma _{j_{1}}\sigma _{j_{2}}\dots \sigma
_{j_{N}}\rangle _{t}$. For our purposes here, we need only the first two
terms: 
$2^{L}\mathcal{P}(\{\sigma ^{\prime }\},t^{\prime }|\{\sigma
\},t)=1+\sum_{k,\ell }\sigma _{k}^{\prime }\sigma _{\ell }M_{k\ell }\left(
t^{\prime }\right) +...\,\,,  \label{P}$
and arrive at 
\begin{equation}
c_{j,k}(t^{\prime };t)=\sum_{\ell }M_{j\ell }\left( t^{\prime }\right)
\langle \sigma _{\ell }\sigma _{k}\rangle _{t}\,\,.  \label{2tcorr}
\end{equation}
The interpretation of this formula is clear: All correlations present at
time $t$ will be ``propagated'' by $M$ over the time delay $t^{\prime }$ and
summed accordingly.

As an illustration, we apply (\ref{2tcorr}) to compute the autocorrelation
function $A_{2k,\ell }(t)\equiv c_{2k,\ell }(t,0)-m_{2k}(t)m_{\ell }(0)$ for
a homogeneous system with initial magnetisation $\bar{m}$ (and $\alpha \neq
0 $). The result is $A_{2k,\ell }(t)=\frac{1-\bar{m}^{2}}{2\sqrt{\gamma _{o}}%
}\left[ \sqrt{\gamma _{e}}+\sqrt{\gamma _{o}}+(-1)^{\ell }(\sqrt{\gamma _{o}}%
-\sqrt{\gamma _{e}})\right] e^{-t}I_{2k-\ell }(\alpha t)$. This simple
example shows that the amplitude of the autocorrelation function alternates
with the parity of site $\ell $ and depends on the temperatures.

\vspace{0.15cm}

\textit{Higher correlation functions:} Finally let us turn to equal-time
correlations of $N$ spins: $\langle \sigma _{j_{1}}\dots \sigma
_{j_{N}}\rangle _{t}$. For the specific case of an initially uncorrelated
system with zero magnetisation, all functions with odd $N$ vanish, of
course. For $N=2n$, we show that they can all be expressed in terms of
two-spin functions (\ref{ETcor}).

This program is achieved by taking advantage of recent formal results
obtained by Aliev \cite{Aliev}. He considered a completely general version
of the kinetic Ising chain, with $1-\tilde{\sigma}_{j}\left( c_{i}\sigma
_{j-1}+d_{i}\sigma _{j+1}\right) /2$ instead of $\left[ 1-\gamma _{j}\tilde{%
\sigma}_{j}\left( \sigma _{j-1}+\sigma _{j+1}\right) /2\right] $ in the
spin-flip rates (\ref{SF}). Such a model would correspond to a system with
not only arbitrary nearest-neighbour couplings ($J_{k,k+1}$), but also a
separate bath (at $T_{k}$) for each spin! In the absence of initial
magnetisation and correlations, Aliev was able to show that the generating
function for all correlations, 
$\Psi (\left\{ \eta \right\} ;t)\equiv \left\langle \prod_{j}\left( 1+\eta
_{j}\sigma _{j}\right) \right\rangle _{t}\,\,,$
where the $\eta $'s are Grassmannian variables \cite{Itz}, is given by 
$\Psi =\mathrm{exp}\left( \sum_{j<k}\eta _{j}\eta _{k}\,\langle \sigma
_{j}\sigma _{k}\rangle _{t}\right)$. Using the notation of \textit{Pfaffians} 
\cite{Itz,MM}, we can expand $\Psi$
and arrive at an expression for $2n$-point function ($j_{1}<j_{2}<...<j_{2n}$%
): 
\begin{equation}
\langle \sigma _{j_{1}}\sigma _{j_{2}}\dots \sigma _{j_{2n-1}}\sigma
_{j_{2n}}\rangle _{t}=\sum_{\pi }\frac{(-1)^{\mathrm{Sg}\pi }}{n!}\,\langle
\sigma _{j_{\pi (1)}}\sigma _{j_{\pi (2)}}\rangle _{t}\dots \langle \sigma
_{j_{\pi (2n-1)}}\sigma _{j_{\pi (2n)}}\rangle _{t}\,\,,  \label{all}
\end{equation}
where the summation runs over all the permutations $\pi $ of the indices $%
\{j_{1},j_{2},...,j_{2n-1},j_{2n}\}$, with the constraint that $j_{\pi
(2\ell -1)}<j_{\pi (2\ell )}$ for each $\ell .$ Here, $\mathrm{Sg}\pi $ is
the signature of the permutation $\pi $. For example, the general
time-dependent 4-spin correlation is given by ($i<j<k<\ell $): 
$\langle \sigma _{i}\sigma _{j}\sigma _{k}\sigma _{\ell }\rangle _{t}=\langle
\sigma _{i}\sigma _{j}\rangle _{t}\langle \sigma _{k}\sigma _{\ell }\rangle
_{t}-\langle \sigma _{i}\sigma _{k}\rangle _{t}\langle \sigma _{j}\sigma
_{\ell }\rangle _{t}+\langle \sigma _{i}\sigma _{\ell }\rangle _{t}\langle
\sigma _{j}\sigma _{k}\rangle _{t}$. In the steady state, the last two terms cancel. More generally, such
cancellations can be shown to persist for arbitrary $n$, so that we recover
the result of Ref.\cite{Beate1}: $ \langle \sigma _{j_{1}}\dots \sigma _{j_{2n}}\rangle _{\infty }=\langle
\sigma _{j_{1}}\sigma _{j_{2}}\rangle _{\infty }\dots \langle \sigma
_{j_{2n-1}}\sigma _{j_{2n}}\rangle _{\infty }$.

Before closing, we remind the readers that the results in this section hold
only for an initially uncorrelated state with zero magnetisation. Otherwise,
the $N$-point functions will clearly be more complex than (\ref{all}).
Finally, due to equation (\ref{rel}), we see that the full distribution $%
P(\{\sigma \},t)$ for this specific nonequilibrium many-body problem can be
constructed from (\ref{ETcor}) and (\ref{all}).

\vspace{0.15cm}

\textit{Concluding remarks:} In this letter we solved a stochastic Ising
chain in which alternate spins are coupled to two thermal baths at different
temperatures via Glauber spin-flip dynamics. We found analytic expressions
for all correlation functions. If both temperatures are positive, both the
steady state and the decays into it display properties similar to those in
the ordinary Glauber-Ising model. If the temperatures are of opposite signs,
qualitatively novel behaviours, such as oscillatory damping, emerge. Similar
to the spatial oscillations in stationary 2-spin correlations, we believe
their origins can be thought of as a kind of ``frustration,'' where the two
baths attempt to align/anti-align a spin with its neighbours.

Finally, we note that our findings can be applied to studies of the dynamics
of \textit{domain walls} in this system \cite{Beate2}. As usual, the kinetic
Ising model can be mapped onto a reaction-diffusion system (RDS),
in which a ``particle'' ($A$) corresponds to a broken bond on the Ising
lattice \cite{Privman}. The resulting RDS, in addition to symmetric
diffusion (with rate $1$), would be pair-annihilation ($AA\rightarrow
\emptyset \emptyset $) with rate $1+\gamma _{j}$ and pair-creation ($
\emptyset \emptyset \rightarrow AA$) with rate $1-\gamma _{j}$. Since $
\left| \gamma _{j}\right| \leq 1$, we are satisfied that these rates are
positive and, so, physical. Our two-temperature model is then mapped into an
RDS with two different creation/annihilation rates on alternating sites. The
implications of mapping our results into the RDS case are interesting, e.g., 
\textit{oscillatory} damping of the density of domain-walls when $\gamma
_{e}\gamma _{o}<0$. Further details will be published elsewhere \cite{prep}.

\vspace{0.2cm}

We are grateful to I.T. Georgiev and U.C. T\"{a}uber for illuminating
discussions. MM acknowledges financial support of Swiss NSF Fellowship N.
81EL-68473. This work was partially supported by US\ NSF DMR-0088451 and
DMR-0308548.

\end{document}